Advisors and indicators based on the SSA models and non-linear generalizations

A.M. Avdeenko

The National Research Technological University, Moscow, Russia

119049, Moscow, Leninsky prospekt, 4

e-mail: aleksei-avdeenk@mail.ru

**Abstract** This paper considers method of creation of an advisor and indicator based on the spectral stochastic analysis model, both with linear and non-linear approximation. The problem of entrance to one or another trade position is solved on the basis of combined analysis of dynamics of quotations of all currency pairs, what allows to actively hedge open positions.

**Keywords**: Forex market, advisor, indicator, singular stochastic analysis

1. Introduction

Great number of models is known, which are used for automation of exchange trade (advisors) and visualization of the recommendations (indicators). These are models of sliding average MA(q), autoregression AR(p), mixed models, such as ARMA(p,q) [1,2], ITRIX, Bolinger bands, Elliott waves, moment and stochastic models, etc.

Models of ARCH(q) type represent conditional dispersion as linear function of squares of past disturbances. In models GARCH(p,q) conditional dispersion at given time is linear function of conditional dispersion and squares of disturbances at past times [3,4]. Further rectification of the GARCH(p,q) model – model EGARCH(p,q) – is able to allow for asymmetry effects – negative correlation between profitability and volatility.

Actual application of these models for creation of automated trading systems or visualization of the tendencies is not too efficient. It frequently happens that the results are trivial – the model's coefficients are either statistically non-significant, or are determined with accuracy, which does not suffice for making of efficient decisions.

In this work, for increase of reliability of forecasts of dynamics of given currency pair, we propose to use information on dynamics of other currency pairs. It will allow to partially avoid effects of non-stability (slow drift of the description parameters in standard models of types MA(q), ARCH(q), etc), and thereby to improve forecasting power and practical efficiency of the model.

## 2. SSA model

Let us assume that input information – is discrete financial sequence $x_n = x(t_n)$ of quotations of a currency pair at time $t_n$. In order to exclude unessential oscillations (lesser than the spread), let us understand $x_n$ as weighted average of prices of opening and closing, and of maximum and minimum for given time frame, which will be assumed equal to minimum possible $\tau = 1$ min. Let us introduce value $y_n = x_{n+1} - x_n$ - change of $x_n$ in the course of evolution of the systems.

Value $y_n$ will be considered as random process. Dozens of currency pairs are simultaneously traded on the market, so let $y_{ni}$ - be relevant change for pair $i = 0 \ldots M - 1$.

Elementary estimations show that currency pairs are correlated, so, for instance, for pairs EURUSD and GBPUSD the correlation factor in averaging interval equal to 20-50 time frames is positive and is within the interval 0.65….0.75, etc. Therefore, when making decision on entrance to / exit from short or long position for given pair, it is expedient to allow for dynamics of changes for the rest currency pairs.

To this end, we offer the following algorithm. Let us introduce information matrix, which reflects dynamics of change of all currency pairs at given time $t_n$

$$U = \begin{pmatrix} y_{n\,0} & y_{n\,1} & \cdots & y_{n\,M-1-K} \\ \cdots & \cdots & \cdots & \cdots \\ y_{n\,K-1} & y_{n\,K} & \cdots & y_{n\,M-1} \end{pmatrix}$$

In this case $M$ - is number of analyzed currency pairs. Now let us determine the two-point correlation function – square matrix with dimensions $(M - K) \times (M - K)$, where value $K$ - is averaging interval in the space of currency pairs at given time: $R_2 = \dfrac{U^T U}{K}$, while $(1 \leq K < M)$.

The matrix is symmetric, therefore eigen values $\lambda$ – that is solutions of equation $det(R_2 - I\lambda) = 0$, are real. To find them, it is convenient to use Jacobi algorithm, which is based on sequence of rotations of the initial matrix, which sequence is implemented in such a way that at each rotation step the value of non-diagonal element, which is maximum by absolute value, is made equal to zero.

Components of the rotation matrix $A^i$ at each step $i$ are $\sin(\theta_{mn}), \cos(\theta_{mn})$, where $\theta_{mn} = \frac{1}{2} arctg \frac{R_{mn}^{i-1}}{R_{mm}^{i-1} - R_{nn}^{i-1}}, R_{mn}^{i-1}$ - is maximum non-diagonal element at the preceding step, if $i-1$.

Maximum number of iterations is $0.5((M-K)^2 - M)$, but actually the procedure may be stopped when modulus of the maximal non-diagonal element becomes less than pre-established value $\varepsilon$. Usually it is assumed that $\varepsilon = 0.001$. Columns of the rotation matrix $A^c = A^1 A^2 ... A^n ...$ – are the initial matrix's eigen vectors, while corresponding eigen values – are diagonal elements of matrix $R = A_c^T R A_c$.

Let us regularize eigen values by absolute value, and let us assign them to eigen vectors. The idea of linear filtration of the forecast in the SSA algorithm consists in retaining of several eigen vectors corresponding to maximum eigen values.

Linear filtration (forecast) in this model is carried out by means of restoration of the process by $l$ eigen values corresponding to maximal, by absolute value, eigen values; in other words, future is forecasted as $y_{n+1 i} = \sum_{q=0}^{l-1} A_{iq}^c y_{nq}$ while $1 < l < M - K$.

In the non-linear filtration model the aforesaid algorithm is implemented for time $t_{n-1}$, and then, by means of the least-squares method, or by means of the reverse-spread neural networks, we build approximation $y_{ni} = \varphi(y_{n-1 i})$.

Now, the future may be represented as $y_{n+1 i} = \varphi(y_{ni})$ and used as a criterion for entrance to short or long position.

Further improvement of the SSA method is possibility to allow for time lag. In this case, instead of the information matrix $U$ we use cellular matrix

$$F = \begin{pmatrix} U_n & ... & ... & U_{n+N-I-1} \\ ... & ... & ... & ... \\ U_{n+I-1} & ... & ... & U_{n+N-1} \end{pmatrix}$$

Its elements are the $U$ matrix's elements shifted for unit time, where $N$ – is total number of shifts, $1 \leq I < N$ – is depth of averaging.

Further analysis and filtration are similar to described above. This not only allows us to take into consideration joint influence of various currency pairs on formation of the exchange rate, but also to allow for lagging with various characteristic time scale.

Practically, for the Forex market the algorithm is implemented in the MetaTrader 5 medium, both as indicators SSA1 – linear filtration and SSA2 – linear filtration with time lag, and one of the advisor's unit as well [8].

Peculiarity of implementation of the algorithm for the advisor is possibility of linear filtration on the basis of the reverse-spread neural network with T hidden layers ($1<T<10$). Method of adaptive behavior described in work [7] is also implemented.

### 3. Testing of the algorithm

Experimental testing of the algorithms SSA1 and SSA2 has been carried out for pair EURUSD for period from 30.09.2012 to 30.09.2013. Currency pairs used for analysis were EURUSD, GBPUSD, USDCHF, USDJPY, USDCAD, AUDUSD, NZDUSD, EURAUD.

Time frame was 1 min, initial deposit $10000, shoulder 100, lot 0.1 of the standard one. Preliminary averaging $p = 5$ was taken into consideration for $l = 1...4$ main components. The algorithm with time lag used neural network with two hidden layers. The algorithm is implemented in the MQL5 medium, and is part of general algorithm [8]. Table 1 presents results of the testing.

Table 1 Results of testing algorithm for the period 1.09.2013-29.09.2013; $l$ - number of eigenvectors, P - profit, Sh-Sharpe ratio and drawdown D%

|  | $l$ | P,$ | Sh | D% |
|---|---|---|---|---|
| SSA1 | 1 | 1822.3 | 0.13 | 6.19 |
|  | 2 | 1277.9 | 0.08 | 7.07 |
|  | 3 | 1216.5 | 0.08 | 7.07 |
|  | 4 | 1216.2 | 0.08 | 7.05 |
| SSA2 | 1 | 1547.5 | 0.08 | 6.64 |
| | 2 | 1530.6 | 0.10 | 6.91 |
| | 3 | 1198.6 | 0.08 | 7.06 |
| | 4 | 1140.3 | 0.08 | 6.95 |

For one-year intervals all implementations appeared to be profitable, while for analysis of two-month intervals the profitability share was 0.65-0.75. On the average, 300-370 trades were made annually. The results appeared to be weakly depending on number of the retained main components, though we observed some decrease of profitability *vs* growth of $l$.

The model with time lag showed no significant advantage as compared with linear filtration, though scattering of the results was somewhat smaller. The sagging was minimal, while Sharp's factor was maximal in the linear model with minimal retaining of the main components.

The SSA algorithm carries out the risk hedging. In the most simple variant the hedging is carried out even for trade of one currency pair, while decision on entrance to / exit from a position is made for this currency pair on the basis of analysis of all currency pairs. When mutual correlations are present, this decreases risk of entrance errors as compared with the situation when decision is made on the grounds of analysis of only one currency pair. More complicated hedging may be implemented by redistribution of the lot value of each currency pair basing on statistics of previous trades and some extreme principles. One of the variants is considered in works [5-8], and is accessible within the package available on the website [9].